\documentclass[fleqn,twoside,twocolumn,nofootinbib]{revtex4} 
\usepackage[sec]{ujp} 
\begin{document}
\title[STATISTICAL DESCRIPTION OF ELECTRODIFFUSION PROCESSES]
{STATISTICAL DESCRIPTION\\ OF ELECTRODIFFUSION PROCESSES\\ IN THE ELECTRON SUBSYSTEM OF A SEMIBOUNDED\\ METAL
WITHIN THE GENERALIZED ``JELLIUM'' MODEL}%
\author{P.P.~Kostrobij}
\affiliation{Lviv Polytechnic National University}
\address{12, Bandera Str., Lviv 79013, Ukraine}
\email{bogdan_markovych@yahoo.com}
\author{B.M.~Markovych}%
\affiliation{Lviv Polytechnic National University}%
\address{12, Bandera Str., Lviv 79013, Ukraine}%
\email{bogdan_markovych@yahoo.com}
\author{A.I.~Vasylenko}
\affiliation{Institute of Physics of Condensed Systems, Nat. Acad. of Sci. of Ukraine}%
\address{1, Sventsytskyi Str., Lviv 79011, Ukraine}%
\author{M.V.~Tokarchuk}%
\affiliation{Institute of Physics of Condensed Systems, Nat. Acad. of Sci. of Ukraine}%
\address{1, Sventsytskyi Str., Lviv 79011, Ukraine}%
\udk{536; 537} \pacs{05.60.Gg, 05.70.Np,\\[-3pt] 63.10.+a, 68.43,
82.20.Xr} \razd{\secix}

\setcounter{page}{179}%
\maketitle

\begin{abstract}
Based on the calculation of the quasiequilibrium statistical sum by
means of the functional integration method, we obtained a
nonequilibrium statistical operator for the electron subsystem of a
semibounded metal in the framework of the generalized ``jellium''
model in the Gaussian and higher approximations with respect to the
dynamic electron correlations. This approach allows one to go beyond
the linear approximation with respect to the gradient of the
electrochemical potential corresponding to weakly nonequilibrium
processes and to obtain generalized transport equations that
describe nonlinear processes.
\end{abstract}

\section{Introduction}

Equilibrium characteristics and nonequilibrium processes of
diffusion, adsorption, and desorption for spatially inhomogeneous
electron-atom systems are described with the help of various
theoretical approaches, both available and being developed ones. In
particular, one widely uses the time-dependent density functional
theory (TDDFT)
\cite{TDF0,TDF1,TDF2,TDF3,TDF6,TDF7,TDF8,TDF11,TDF12,TDF14,TDF15,TDF18,TDF20}.
During the years of its development, the TDDFT has demonstrated
significant achievements and still extends its limits of application
\cite{TDF20} though with certain problems \cite{TDF23}. The basis of
the TDDFT is the Kohn--Sham density functional theory
\cite{l37,l38,l38a,l40,l41}. Another theoretical approach is related
to the hydrodynamic model of surface plasmons for a
spatially inhomogeneous electron gas proposed in \cite{EQ1,EQ2,EQ3}
with the use of the response theory \cite{Grif} based on the
Boltzmann kinetic equation. The quantum statistical theory for
the description of nonequilibrium processes in ``metal--adsorbate--gas''
systems was developed in works \cite{Kost5,Kost111,Kost000} using
Zubarev’s method of nonequilibrium statistical operator (NSO)
\cite{l121,l122}. In particular, a self-consistent description of
nonequilibrium processes in the atomic and electron subsystems
was presented in \cite{Kost5} on the
kinetic level of the description of electron processes. In the processes
of adsorption, desorption, and surface diffusion, a metal surface
undergoes a reconstruction accompanied by a variation of
nonequilibrium properties of both electron and ion subsystems. In
this case, the electrodiffusion, viscothermal, and electromagnetic
properties of the electron subsystem change in the field of metal
surface ions. To study the ion and electron structures of a
semibounded metal, a generalized approach
that takes the effect of
discreteness of the ion subsystem into account and is based upon the model of
semibounded ``jellium'' \cite{l184,l185} was proposed in \cite{Kost111,Kost000}. It is worth noting that
the influence of the discreteness of an ion density on the
characteristics of a semibounded ``jellium'' was considered in
\cite{l41,sss1,sss2,sss3} by means of constructing a perturbation
theory with respect to the electron-ion interaction pseudopotential.
However, the linear response of the electron subsystem to the
lattice potential did not take the effects of
inhomogeneity of the electron subsystem into account. The approach described in
\cite{l184,l185, Kost111} allows one to model the formation of a
surface potential and to calculate a large statistical sum for the
generalized model in terms of the cumulant averages of the
``jellium'' model. In \cite{Kost111}, the generalized ``jellium''
model served as a basis for the statistical description of
electrodiffusion processes for the electron subsystem of a
semibounded metal with the use of the NSO method, where the only
parameter of the reduced description was the nonequilibrium average
value of the electron density. For such a system, the
quasiequilibrium statistical sum was calculated by means of the
functional integration method for the case of the local electron-ion
interaction pseudopotential of the metal surface. In principle, it
allowed one to obtain expressions for the nonequilibrium statistical
operator in the Gaussian and higher approximations with respect to
the dynamic electron correlations. In \cite{Kost111}, the
nonequilibrium statistical operator and the generalized transport
equation of inhomogeneous diffusion were obtained for weakly
nonequilibrium processes (linear approximation with respect to the
gradient of the electrochemical potential). The same approximation
was used to deduce an equation for the ``density-density'' time
correlation function that determines the dynamic structural factor
of the electron subsystem of a semibounded metal and to demonstrate
the connection of this electrodiffusion model in the linear
approximation with the TDDFT \cite{TDF0,TDF1,TDF2,TDF3}.

The given study represents the continuation of work \cite{Kost111}.
We will obtain expressions for the nonequilibrium statistical operator
in the Gaussian and higher approximations with respect to the
dynamic electron correlations, by calculating the quasiequilibrium
statistical sum by means of the functional integration method. This
approach allowed us to go beyond the linear approximation with
respect to the electrochemical potential. For the nonequilibrium
statistical operator in the corresponding approximations, we
will obtain generalized transport equations for the nonequilibrium
average value of the electron density for strongly nonequilibrium
processes for the electron subsystem of a semibounded metal.

\section{Generalized ``Jellium'' Model. Nonequilibrium Statistical Operator}

\subsection{Hamiltonian of the system}

Consider an electron-ion system that describes a semibounded metal
with regard for the influence of the discreteness of the ion
subsystem. We present the Hamiltonian of the system in the form
\[H = -\frac{\hbar^2}{2m}\sum\limits_{i=1}^{N}\Delta_i+
                  \frac12\sum\limits_{i\neq j=1}^{N}
                  \frac{e^2}{|\mathbf{r}_i-\mathbf{r}_j|} +\]
\begin{equation}\label{Ham1}
   +\frac12\sum\limits_{i\neq j=1}^{N_{\rm ion}}
                \frac{({\mathcal{Z}} e)^2}{|\mathbf{R}_i-\mathbf{R}_j|}
                +\sum\limits_{i=1}^N\sum\limits_{j=1}^{N_{\rm ion}}
                e\,w(\mathbf{r}_i,\mathbf{R}_j),
 \end{equation}
where the first two terms represent the electron kinetic energy
and the potential energy of electron-electron interaction,
respectively, the third term stands for the potential energy of
ion-ion interaction, and the last one is the energy of electron-ion
interaction. Electrons in the ion field have the charge $e$, the
mass $m$, and the coordinates $\mathbf{r}_i$, $i=1,\ldots,N$.
By $N_{\rm ion},$ we denote the number of metal ions with the charge
${\mathcal{Z}} e$ and the coordinates $\mathbf{R}_j$
($-\infty<X_j,Y_j<+\infty$,
 $Z_j\leqslant z_0$, $Z_0={\rm const}$, $z=Z_0$ is the division plane), $j=1,\ldots,N_{\rm ion}$.
We suppose that ions are immobile in the system volume $V=SL$, where $S$ is the surface area of the
semibounded metal, and $L$ determines the region of variation of the electron coordinate normal
to the metal surface: ${z\in(-L/2,+L/2)}$, $S\to\infty$, $L\to\infty$. We consider that the system is electroneutral, i.e.
 \begin{equation}\label{Neytral}
    {\mathcal{Z}} N_{\rm ion}=N.
 \end{equation}

In \cite{Kost111}, Hamiltonian (\ref{Ham1}) was presented in terms of the collective variables of
the electron subsystem of a semibounded metal specifying the Hamiltonian of the ``jellium'' model as a reference system:
\[H=\sum\limits_{\mathbf{p},\alpha}E_\alpha(\mathbf{p})
   a^\dagger_{\alpha}(\mathbf{p})
   a^{\vphantom\dagger}_{\alpha}(\mathbf{p}) +\]
\[+ \frac{1}{2SL}\sum\limits_{\mathbf{q}}{}^{'}
  \sum\limits_k
  \nu_k(\mathbf{q})\rho_k(\mathbf{q})\rho_{-k}(-\mathbf{q})-\]
\[-\frac{{\mathcal{Z}} N_{\rm ion}}{SL}
  \sum\limits_{\mathbf{q}}{}^{'}\sum\limits_k
  \nu_k(\mathbf{q})\mathrm{S}_k(\mathbf{q})\rho_k(\mathbf{q}) +\]
\[+\frac{e N_{\rm ion}}{SL}
  \sum\limits_{\mathbf{q},k}\mathrm{S}_k(\mathbf{q})
  f_{k}(\mathbf{q})\rho_{k}(\mathbf{q})-\frac{N}{2S}\sum\limits_{\mathbf{q}}{}^{'}\nu(\mathbf{q}|0)
  +\]
  \begin{equation}\label{Ham4}
+\frac12\sum_{i\neq
  j=1}^{N_{\rm ion}}\frac1S\sum_\mathbf{q}{}^{'}\!\!
  {\mathcal{Z}}^2\nu(\mathbf{q}|Z_i-Z_j){e}^{{i} \mathbf{q}(\mathbf{R}_{||i}-\mathbf{R}_{||j})},
 \end{equation}
 where the primed sums mean the absence of terms with $\mathbf{q}=0$ due to the electroneutrality
condition (\ref{Neytral}), $\nu_k(\mathbf{q})=4\pi e^2/(q^2+k^2)$ and $f_k(\mathbf{q})$ are
the three-dimensional Fourier transforms of the Coulomb potential and the local part of pseudopotential (\ref{ModPot}):
  \[
  \frac{e^2}{|\mathbf{r}_i-\mathbf{r}_j|}=\frac1{SL}
  \sum_{\mathbf{q},k}\nu_k(\mathbf{q})
  {e}^{{i} \mathbf{q}(\mathbf{r}_{||i}-\mathbf{r}_{||j})+
  {i} k(z_i-z_j)},
 \]
  \begin{equation}\label{ModPot}
  w(\mathbf{r}_i-\mathbf{R}_j)=-\frac{{\mathcal{Z}} e}{|\mathbf{r}_i-\mathbf{R}_j|}
  + f(\mathbf{r}_i-\mathbf{R}_j),
 \end{equation}
 \[
  f(\mathbf{r}_i-\mathbf{R}_j)=\frac1{SL}
  \sum_{\mathbf{q},k}f_k(\mathbf{q})
  {e}^{{i} \mathbf{q}(\mathbf{r}_{\parallel i}-\mathbf{R}_{\parallel j})+
  {i} k(z_i-Z_j)},
 \]
 $
 \mathbf{R}_{\parallel j}=(X_j,Y_j),
 $
 $\nu(\mathbf{q}|z)=2\pi e^2 {e}^{-q|z|}/q$ is the two-dimensional Fourier transform of the Coulomb potential: $\dfrac{e^2}{r}=\dfrac1S
  \sum\limits_\mathbf{q}\nu(\mathbf{q}|z){e}^{-\mathbf{q}\mathbf{r}_{\parallel}},$
  $
  E_\alpha(\mathbf{p})=\dfrac{\hbar^2p^2}{2m}+\varepsilon_\alpha $
 is the energy of an electron in the state $(\mathbf{p},\alpha)$,
 \begin{equation}\label{StrFaktor}
   \mathrm{S}_k(\mathbf{q})=\frac1{N_{\rm ion}}
   \sum\limits_{j=1}^{N_{\rm ion}}{e}^{-{i}\mathbf{q}\mathbf{R}_{\parallel j}-{i} k
   Z_j}
 \end{equation}
  stands for the structural factor of the ion subsystem and the Fourier transform of the electron density:
 \begin{equation}\label{ro}
   \rho_k(\mathbf{q})=\sum\limits_{\mathbf{p},\alpha_1,\alpha_2}
   \langle\alpha_1|{e}^{{i} k z}|\alpha_2\rangle
   a^\dagger_{\alpha_1}(\mathbf{p})
   a^{\vphantom\dagger}_{\alpha_2}(\mathbf{p}-\mathbf{q}),
 \end{equation}
  where
 $\langle\alpha_1|\ldots|\alpha_2\rangle=\displaystyle\int\limits\!{\rm d}
  z\, \varphi^*_{\alpha_1}(z)\ldots\varphi_{\alpha_2}(z).$
$\varphi_\alpha(z)$ and $\varepsilon_\alpha$ are the eigenfunctions and eigenvalues of the Sch\"odinger equation, respectively,
 \[
    \left[-\frac{\hbar^2}{2m}\frac{{\rm d}^2}{{\rm d} z^2}+V(z) \right]
    \varphi_\alpha(z)=\varepsilon_\alpha\varphi_\alpha(z),
 \]
   $V(\mathbf{r})=V(z)$ is the surface potential depending only on the electron coordinate normal to the division plane.

The same way as in \cite{Kost111}, the electrodiffusion processes in the formulated model are described choosing
the average value of the electron density operator as the main parameter of the reduced description
of nonequilibrium processes in the electron subsystem of a semibounded metal. It is connected with
the corresponding inhomogeneous electric field,
\begin{equation}\label{n1}
{\bf \nabla}\cdot {\bf E}({\bf r};t)=e \langle \varrho({\bf
r})\rangle^{t},
 \end{equation}
where $\langle(...)\rangle^{t}= {\rm Sp}(...) \rho(t)$, $\rho(t)$ denotes the nonequilibrium statistical
operator of the generalized ``jellium'' model that satisfies the Liouville equation with the
Hamilton operator (\ref{Ham4}). With regard for the chosen geometry of the model, the value
of $\langle \varrho({\bf r})\rangle^{t}$ will correspond to the mixed Fourier representation
$\langle \rho_{k}({\bf q})\rangle^{t}$. To find $\rho(t)$ (the solution of the Liouville equation),
we employ Zubarev’s NSO method \cite{l121,l122} and obtain, in the general case,
 \begin{equation} \label{1.23}
\rho(t) = \varepsilon \int\limits_{-\infty}^t {\rm
e}^{\varepsilon(t'-t)}{e}^{iL_N(t'-t)}\rho_q(t'){\rm d}t',
\end{equation}
where $\varepsilon\to+0$ after passing to the thermodynamic limit, and $iL_N$ is the Liouville operator
corresponding to Hamilton operator (\ref{Ham4}). The quantity $\rho_{q}(t)$ denotes the quasiequilibrium
statistical operator determined by the Gibbs method at fixed values of the parameter of the
reduced description $\langle \rho_{k}({\bf q})\rangle^{t},$ and the normalization condition
${\rm Sp} \rho_{q}(t)=1$ hold. In our case, it has the following form \cite{Kost111}:
\[\rho_{q}(t)= \exp \Bigg[-\Phi(t) - \]
\begin{equation}\label{n2}
 -\beta \left(H- \frac {1}{SL}\sum_{k}\sum_
{{\bf q}}{\tilde \mu}_{k}({\bf q};t)\rho_{k}({\bf q})\right)\Bigg] ,
\end{equation}
where $\Phi(t)=\ln Z(t)$ is the Massieu--Planck functional, and $Z(t)$
is the statistical sum of the quasiequilibrium statistical
operator,
\begin{equation}\label{n3}
Z(t)={\rm Sp}\exp  \Bigg[- \beta \left(H- \frac {1}{SL}\sum_{k}\sum_ {{\bf
q}}{\tilde \mu}_{k}({\bf q};t)\rho_{k}({\bf q})\right)\Bigg],
\end{equation}
${\tilde \mu}_{k}({\bf q};t)=\mu_{k}({\bf q};t)+e\varphi_{k}({\bf
q};t)$ denotes the Fourier transform of the electron electrochemical potential,
$\mu_{k}({\bf q};t)$ is the Fourier transform of the electron chemical potential, and $\varphi_{k}({\bf q};t)$
is the Fourier transform of the local electric potential. The quantity ${\tilde \mu}_{k}({\bf q};t)$ is determined
from the self-consistency condition
\begin{equation}\label{n4}
\langle \rho_{k}({\bf q})\rangle^{t}=\langle\rho_{k}({\bf
q})\rangle_{q}^{t}
\end{equation}
and the thermodynamic relations
\begin{equation}\label{n4*}
\frac{\delta\Phi(t)}{\delta\frac{\beta}{SL}\bar{\mu}_k({\bf
q};t)}=\langle\rho_k({\bf  q})\rangle^t,
\end{equation}
\begin{equation}\label{n5}
\begin{array}{l}
  \displaystyle\frac {{\delta S(t)}}{{\delta \langle \rho_{k}({\bf
q})\rangle^{t}}}=-\frac{\beta}{SL}\mu_{k}({\bf q};t),  \\[8mm]
 \displaystyle \frac {{\delta S(t)}}{{\delta \langle e\rho_{k}({\bf
q})\rangle^{t}}}=-\frac{\beta}{SL}\varphi_{k}({\bf q};t),
\end{array}
\end{equation}
where $S(t)$ is the Gibbs nonequilibrium entropy,
\[
S(t)=-{\rm Sp}{(\ln \rho_{q}(t))\rho_{q}(t)}= \]
 \[=\Phi(t)+\beta \Big(\langle H
\rangle^{t}-\frac{1}{SL}\sum_{k}\sum_ {{\bf q}}{\tilde \mu}_{k}({\bf
q};t)\langle \rho_{k}({\bf q})\rangle^{t}\Big)=\]
\[
=\ln Z(t)+\beta \Big(\langle H \rangle^{t}-\frac{1}{SL}\sum_{k}\sum_
{{\bf q}} \mu_{k}\Big({\bf q};t)\langle \rho_{k}({\bf
q})\rangle^{t}-\]
\begin{equation}\label{n7}
-\frac{1}{SL}\sum_{k}\sum_ {{\bf q}} \varphi_{k}({\bf q};t)\langle e
\rho_{k}({\bf q})\rangle^{t}\Big),
\end{equation}
$\langle e \rho_{k}({\bf q})\rangle^{t}=e\langle \rho_{k}({\bf
q})\rangle^{t}$ is the average electron charge density. As follows from the structure
of the nonequilibrium entropy, the transport processes in the system within the used model are caused by
the gradients of the local chemical and electrochemical potentials.

With regard for the structure of $\rho_{q}(t)$, the nonequilibrium statistical operator can be presented in the form
\[
\rho(t) = \varepsilon \int\limits_{-\infty}^t
e^{\varepsilon(t'-t)}e^{iL_N(t'-t)} e^{-\hat{S}(t')}{\rm d}t'=\]
\[=\rho_{q}(t)+\int\limits_{-\infty}^t
e^{\varepsilon(t'-t)}e^{iL_N(t'-t)}\times\]
\begin{equation} \label{1.23a}
\times\int\limits_{0}^{1}\!{\rm d}\,\tau\rho_{q}^{\tau}(t')
\left(\frac{\partial}{\partial
t'}+iL_N\right)\hat{S}(t')\rho_{q}^{1-\tau}(t'){\rm d}t',
\end{equation}
where
\begin{equation} \label{1.23aa}
\hat{S}(t')=\ln Z(t)+\beta\!\!\left(\!H-\frac {1}{SL}\sum_{k}\sum_ {{\bf
q}}{\tilde \mu}_{k}({\bf q};t)\rho_{k}({\bf q})\!\right)
\end{equation}
is the entropy operator.
In order to reveal the structure of the entropy operator, it is necessary to calculate
the statistical sum $Z(t)$ of the quasiequilibrium statistical operator. With regard for the
structure of Hamiltonian (\ref{Ham4}), $Z(t)$  can be put down as follows:
\cite{Kost111}:
\[
 Z(t) = {\rm Sp}\left\{\exp ( - \beta (H_0 - \dfrac{1}{2S}\sum\limits_{\bf  q} {}^{'}\nu
({\bf  q}\left| 0 \right.) +\right. \]
\[ +\dfrac{1}{2SL}\sum\limits_{k,{\bf  q}} \nu _k ({\bf  q})\rho _k
({\bf  q})\rho _{ - k} ( - {\bf  q}) + \]
\begin{equation}\label{n10}
+ \left.\dfrac{1}{2SL}\sum\limits_{k,{\bf  q}} B({\bf  q},k;t)\rho
_{ - k} ( - {\bf  q}))\right\} ,
\end{equation}
  where
  $B({\bf  q},k;t) =
  N_{\rm ion} S_{k} ({\bf  q})\omega_{k}
  ({\bf  q}) - \bar {\mu }_{k} ({\bf  q};t)$, $w_k(\mathbf{q})=
      -{\mathcal{Z}}\nu_k(\mathbf{q})+
      ef_k(\mathbf{q})$,
$
  H_0'=\sum_{\mathbf{p},\alpha} E_\alpha(\mathbf{p})
   a^\dagger_{\alpha}(\mathbf{p})
   a^{\vphantom\dagger}_{\alpha}(\mathbf{p})
$
is the kinetic part of the Hamiltonian of the electron subsystem.

Applying the functional integration method and considering the ``jellium'' model as a reference system,
$Z(t)$ can be written down as \cite{Kost111}
\begin{equation}\label{n11}
 Z(t) = \exp \left\{\beta
\frac{N}{2S}\sum\limits_{{\bf  q}}{}^{'}\nu ({\bf  q}|0 )\right\}
 Z_{{\rm jell}} \, \Delta Z(t),
\end{equation}
where
\begin{equation}\label{n12}
 Z_{{\rm jell}} = {\rm Sp} \big\{\exp (- \beta H_{0}){\rm T}S_1(\beta)\big\}
\end{equation}
is the statistical sum of the ``jellium'' model of the electron subsystem of a semibounded metal
that corresponds to the equilibrium state calculated in \cite{l184,l185};
\[
  S_1(\beta)=\exp\Bigg[-\frac1{2SL}\int\limits_0^\beta\!{\rm d}\beta'
   \sum\limits_{\mathbf{q}}{}^{'}\sum\limits_k
   \nu_k(\mathbf{q})\times \]
   \begin{equation}\label{S1}
   \times\rho_k(\mathbf{q}|\beta')\rho_{-k}(-\mathbf{q}|\beta')\Bigg]
 \end{equation}
is the contribution of the electron interaction, where
$\rho_{k}(\mathbf{q}|\beta')={\rm
e}^{\beta'H_0}\rho_{k}(\mathbf{q}){e}^{-\beta'H_0}$,
\[
 \Delta Z(t) = \frac{1}{Z_{{\rm jell}} }{\rm Sp}\Big\{\exp ( -
\beta H_{0} ){\rm T}S_1(\beta)S_{2}(\beta ;t)\Big\} = \]
\begin{equation}\label{n13}
= \langle S_{2}(\beta ;t)\rangle_{{\rm jell}},
\end{equation}
where $\langle (.....)\rangle_{{\rm
jell}}=\displaystyle\frac{1}{Z_{{\rm jell}} }{\rm Sp}\Big\{\exp ( -
\beta H_{0} ){\rm T}S_1(\beta)(...)\Big\}$;
\[
S_{2}(\beta ;t) = \]
\begin{equation}\label{n14}
 ={\rm T}\exp \,  \Biggl\{
 - \int\limits_0^\beta {\rm d}\beta'
 \frac{1}{SL}\sum\limits_{k,{\bf  q}} B({\bf  q},k;t)\rho _{ - k}
 ( - {\bf  q};\beta ')\!\Biggr\}.
 \end{equation}
Using the cumulant representation, $\Delta Z(t)$ can be presented in
the form
\[
  \Delta Z(t) = \exp \Biggl[\sum\limits_{n=1}
\frac{{i}^{n}}{n!}\left(\frac{\beta
}{SL}\right)^{n}\sum\limits_{{\bf  q}_{1} ...{\bf  q}_{n} }
\sum\limits_{k_{1} ...k_{n} } B({\bf  q}_{1} ,k_{1} ;t)...\times\]
\begin{equation}\label{n16}
\times B({\bf  q}_{n} ,k_{n} ;t)
 {\mathfrak M}_{- k_{1} .... - k_{n} } (- {\bf  q}_{1} .... - {\bf  q}_{n})\Biggr ],
\end{equation}
  where
\[
 {\mathfrak M}_{k_{1} ....k_{n} } ({\bf  q}_{1} ....{\bf  q}_{n} )=
\]
\begin{equation}\label{n17}
 = {i}^{n}\langle {\rm T}\rho
_{k_{1}} ({\bf  q}_{1}| 0 ),....\rho_{k_{n}} ({\bf  q}_{n}|0
)\rangle^{c}_{{\rm jell}}
\end{equation}
 are the cumulant irreducible average values of the electron density fluctuations calculated
with the help of the equilibrium statistical operator of the ``jellium'' model of the electron
subsystem of a semibounded metal \cite{l184,l185}. In particular, the second cumulant has the structure
\[
 {\mathfrak M}_{k_{1} ,k_{2} } ({\bf  q}_{1} ,{\bf  q}_{2} ) = \langle
 \rho_{k_{1}}({\bf  q}_{1})\rho_{k_{2} } ({\bf  q}_{2} )\rangle_{{\rm jell}}  -
\]
\begin{equation}\label{n18}
-\langle \rho_{k_{1}} ({\bf  q}_{1})\rangle_{{\rm jell}} \langle
\rho_{k_{2}} ({\bf  q}_{2})\rangle_{{\rm jell}}
\end{equation}
and is connected with the static structural factor $S(k_{1} , {\bf
q}_{1} ;k_{2} ,{\bf  q}_{2} ) = \langle \rho_{k_{1} } ({\bf q}_{1}
)\rho _{k_{2}} ({\bf  q}_{2})\rangle _{{\rm jell}} $ of the electron
subsystem of a semibounded metal. In the Gaussian approximation, we
obtain
\[
 \Delta Z^G(t) = \exp \Biggl[ - \frac{1}{2}\left(\frac{\beta
 }{SL}\right)^{2}\sum\limits_{{\bf q}_{1} {\bf  q}_{2}} \sum\limits_{k_{1} k_{2} } B({\bf  q}_{1}
,k_{1} ;t) \times\]
\begin{equation}\label{n19}
\times B({\bf  q}_{2},k_{2} ;t){\mathfrak M}_{- k_{1} , - k_{2} } (
- {\bf  q}_{1}, - {\bf  q}_{2} )\Biggr]
\end{equation}
is expressed in terms of the second cumulant of the ``jellium'' model of inhomogeneous electron gas \cite{l184,l185}.
According to the definition of $s$-particle electron distribution functions \cite{l184,l185,Kost111},
we obtain the quasiequilibrium $s$-particle electron distribution functions in the form
\[
  F_s({\bf r}_1,\ldots,{\bf r}_n;t)=F_s({\bf r}_1,\ldots,{\bf r}_n)^{{\rm jell}}
 \times\]
 \[
  \times\exp\Bigg[\sum_{n\geqslant1}\frac{{i}^n}{n!}
  \left(\frac{\beta N_{\rm ion}}{SL}\right)^n
  \!\!\!\!\sum_{{\bf q}_1,\ldots,{\bf
q}_n}{}^{\!\!\!\!\!'}\sum_{k_1,\ldots,k_n} B({\bf  q}_{1} ,k_{1}
;t)\ldots   \times\]
\begin{equation}\label{n20}
\times B({\bf  q}_{n},k_{n} ;t) \Delta{\mathfrak
M}^{(s)}_{-k_1,\ldots,-k_n}(-{\bf q}_1,\ldots,-{\bf q}_n)
  ],
\end{equation}
where
\[
\Delta{\mathfrak M}^{(s)}_{-k_1,\ldots,-k_n}(-{\bf q}_1,\ldots,-{\bf
q}_n) = \]
\[ ={\mathfrak M}^{(s)}_{-k_1,\ldots,-k_n}(-{\bf q}_1,\ldots,-{\bf q}_n)-
\]
\[ -{\mathfrak M}_{-k_1,\ldots,-k_n}(-{\bf q}_1,\ldots,-{\bf q}_n).
\]

Relations (\ref{n20}) link the quasiequilibrium distribution
functions with the electrochemical potential $\bar {\mu }_{k} ( {\bf
q};t)$ through the corresponding cumulant averages of the
``jellium'' model. In view of the structure of $\Delta
Z(t)$ (\ref{n19}), $\ln Z(t)$ can be written as
\[
\ln Z(t)= \beta\frac{N}{2S}\sum_{{\bf  q}}{}^{'}\nu({\bf  q}|0)+\ln
Z_{{\rm jell}}+\]
\[+\sum\limits_{n=1} \frac{{
i}^{n}}{n!}\left(\frac{\beta }{SL}\right)^{n}\sum\limits_{ {\bf
q}_{1} ...{\bf  q}_{n} } \sum\limits_{k_{1} ...k_{n} } B( {\bf
q}_{1} ,k_{1} ;t)...\times\]
\begin{equation} \label{551}
 \times B({\bf  q}_{n} ,k_{n} ;t)
 {\mathfrak M}_{- k_{1} .... - k_{n} } (- {\bf  q}_{1} .... - {\bf  q}_{n}),
\end{equation}
where $\ln Z_{{\rm jell}}$ can be calculated in various approximations with respect to the
electron correlations \cite{l184,l185}. Based on (\ref{551}) and (\ref{1.23aa}), we obtain the expression
for the nonequilibrium statistical operator in the general form:
\[
\rho(t) =\rho_{q}(t)+\int\limits_{-\infty}^t\!\!{\rm d}t' {
e}^{\varepsilon(t'-t)}e^{iL_N(t'-t)} \Bigg\{\frac{{
i}^{n}}{n!}\left(\frac{\beta}{SL}\right)^{n} \!\!\times \]
\[\times
 \,
 \sum\limits_{{\bf  q}_{1} ...{\bf  q}_{n} }
\sum\limits_{k_{1} ...k_{n} } \frac {\partial }{\partial
t'}\Big(B({\bf  q}_{1} ,k_{1} ;t')... B({\bf  q}_{n} ,k_{n}
;t')\Big)\times \]
\[ \times
 {\mathfrak M}_{- k_{1} .... - k_{n} } (- {\bf  q}_{1} .... - {\bf  q}_{n})
 \Bigg\}
\rho_{q}(t')- \]
 \[-\frac {1}{SL}\sum_{k}\sum_ {{\bf q}}
 \int\limits_{-\infty}^t \!\!{e}^{\varepsilon(t'-t)}{e}^{{
 i}L_N(t'-t)}\]
 \begin{equation} \label{1.23abb}
\times\int\limits_{0}^{1}\!{\rm d}\tau\rho_{q}^{\tau}(t')
\dot{\rho}_{k}({\bf q}) \rho_{q}^{1-\tau}(t'){\tilde \mu}_{k}({\bf
q};t'){\rm d}t',
\end{equation}
where $\dot{\rho}_{k}({\bf q})={i}L_N \rho_{k}({\bf q})=-k{\bf
q}\cdot {\bf  J}_k({\bf  q}),$ and ${\bf  J}_k({\bf  q})$ is the
Fourier transform of the microscopic electron flux. The obtained
expression represents a sum of the non-dissipative and dissipative
parts. The first one corresponds to the operator $\rho_q(t)$,
while the second one is described by the terms that contain the
time derivatives of the functions $B({\bf  q},k;t')$ and the
microscopic fluxes $\dot\rho_k({\bf  q})$. Moreover, the
derivative $\frac{\partial}{\partial t'}B({\bf  q},k;t')$ can be
presented as
\[\frac{\partial}{\partial t'}B({\bf  q},k;t')
=\frac{\partial}{\partial t'}\Big(N_{\rm ion} S_k({\bf
q})\omega_k({\bf q}\,)-\tilde{\mu}_k({\bf  q};t')\Big)=\]
\[=-\frac{\partial}{\partial
t'}\tilde{\mu}_k({\bf  q};t')=-\frac{\delta\tilde{\mu}_k({\bf
q};t')} {\delta\langle\rho_k({\bf  q})\rangle^{t'}}
\frac{\partial}{\partial t'}\langle\rho_k({\bf  q})\rangle^{t'},
\]
where
 \begin{equation}\label{v30}
  \frac{\delta\tilde{\mu}_k({\bf  q};t')}
  {\delta\langle\rho_k({\bf  q})\rangle^{t'}}
  = -\left(\rho_k({\bf  q})|\rho_{-k}(-{\bf  q})\right)^{-1}_q,
\end{equation}
 \[\big(\rho_k({\bf  q})|\rho_{-k}(-{\bf  q})\big)^{-1}_q =\int\limits_0^1\!\!\Big\langle\left(\rho_k({\bf  q})-\langle\rho_k({\bf  q})\rangle_q^{t'}\right)\times
 \]
 \begin{equation}\label{v31}
\times \left(\rho_{-k}({-{\bf q}};\tau) -
 \langle\rho_{-k}({-{\bf q}}\,)\rangle^{t'}_q\right)\Big\rangle_q^{t'}{\rm d}\tau,
\end{equation}
is the quantum quasiequilibrium correlation function,
 \[
 \rho_k({\bf  q};\tau)=\rho^{\tau}_q(t')\rho_k({\bf  q})\rho_q^{-\tau}(t').
 \]
With regard for (\ref{v30}), the time derivative of $B({\bf
q},k;t')$ can be presented in the form
\[ \frac{\partial}{\partial t'}B({\bf  q},k;t')=
 \left(\rho_k({\bf  q})|\rho_{-k}(-{\bf  q})\right)^{-1}_q
 \frac{\partial}{\partial t'} \langle\rho_k({\bf  q})\rangle^t=\]
 \begin{equation}\label{v32}
 =\left(\rho_k({\bf  q})|\rho_{-k}(-{\bf  q})\right)^{-1}_q\langle\dot\rho_k({\bf  q})\rangle^t.
\end{equation}

Thus, the nonequilibrium statistical operator (\ref{1.23abb}) is a
functional of the pair quasiequilibrium correlation functions
(\ref{v31}), equilibrium correlation functions (\ref{n17}),
average values of parameters of the reduced description
$\langle\rho_k({\bf  q})\rangle^t$, and microscopic fluxes
$\dot\rho({\bf  q})$ of the electron subsystem of a semibounded
metal. To make the description complete, it is necessary to obtain
the transport equation for $\langle\rho_k({\bf  q})\rangle^t$ with
the help of the non-equilibrium statistical operator (\ref{1.23abb}).
Based on the structure of $\rho(t)$, one can state that these
equations will be nonlinear. The parameters $\bar{\mu}_k({\bf q};t)$
in these equations should be found with the use of the self-consistency conditions
(\ref{n4}). The obtained expression for the quasiequilibrium
statistical operator with the Massieu--Planck functional (\ref{551})
allows one to find the nonequilibrium statistical operator in the
corresponding approximations, particularly in the Gaussian one.

\section{Gaussian Approximation}

Here, we will consider approximation (\ref{n19}), in
which the nonequilibrium statistical operator and the transport
equation for $\langle\rho_k({\bf  q})\rangle^t$ will be obtained.
The index $G$ in all cases means the description of a
function in the Gaussian approximation.

With regard for the structure of (\ref{n19}), we obtain the entropy operator (\ref{n7}) in the form
\[
\hat S^{(G)}(t)=\beta\frac{N}{2S}\sum_{{\bf  q}}{}^{'}\nu( {\bf
q}|0)+\ln Z_{{\rm jell}}-\]
\[-\frac{1}{2}\left(\frac{\beta}{SL}\right)^2\!\!\!
  \sum_{k_1,k_2}\sum_{{\bf  q}_1,{\bf  q}_2}\!\!
  \Big(N_{\rm ion}S_{k_1}({\bf  q}_1)\omega_{k_1}({\bf  q}_2)-\bar\mu_{k_1}({\bf  q}_1;t)\Big)
  \!\!\times\]
\[\times\Big(N_{\rm ion}S_{k_2}(
{\bf q}_2)\omega_{k_2}({\bf  q}_2)
 -\bar\mu_{k_2}({\bf  q}_2;t)\Big)\times\]
 \[
\times{\mathfrak M}_{-k_1,-k_2}
 (-{\bf  q}_1,-{\bf  q}_2)+\]
 \begin{equation} \label{51}
 + \beta\!\Bigg(\!H-\frac{1}{SL}\sum_{k,
{\bf q}}\bar\mu_k({\bf  q};t)\rho_k({\bf  q})\!\Bigg).
\end{equation}

In order to eliminate the parameters $\bar\mu_k({\bf  q};t)$ from
this formula, we use the thermodynamic relation (\ref{n4*})
\[
 \frac{\delta\Phi^{(G)}(t)}{\delta\frac{\beta}{SL}\bar\mu_k({\bf  q};t)}=\langle\rho_k({\bf  q})\rangle^t
\]
which yields
\[
{\langle\rho_k({\bf
q}\,)\rangle^t=\frac{\beta}{SL}}\times\]
\begin{equation} \label{52}
\times\sum_{k',{\bf  q}}\left({\bar S}_{k'}(  {\bf q}')
 -\bar\mu_{k'}({\bf  q};t)\right){\mathfrak M}_{-k',-k}(-{\bf  q},-{\bf  q}){\bar S}_{k}
 ({\bf  q}),
\end{equation}
where
\[
{\bar S}_{k}({\bf  q})=N_{\rm ion}S_{k}({\bf  q})\omega_{k}({\bf
q}).
\]
Defining the function ${\mathfrak M}_{-k,-k'}^{-1}(-{\bf  q},-{\bf
q}')$ inverse to ${\mathfrak M}_{-k,-k'}(-{\bf  q},-{\bf  q})$ by
the relation
\[
\sum_{k'',{\bf  q}'}{\mathfrak M}_{k,k''}^{-1}({\bf  q}, {\bf
q}''){\mathfrak M}_{k'',k'}({\bf  q}'',{\bf
q}')=\delta_{k,k'}\delta_{{\bf  q},{\bf  q}'}
\]
and using (\ref {52}), we obtain the Fourier transform of the electron electrochemical potential as follows:
\[{\bar\mu_k({\bf  q};t)={\bar S}_{k}({\bf  q})}-\left(\frac{\beta}{SL}\right)^{-1}\times\]
\begin{equation}\label{53}
\times\sum_{k',{\bf q}'}\langle\rho_{k'}({\bf  q}')\rangle^t{\bar
S}_{k'}^{-1}({\bf q}')\mathfrak M_{-k',-k}^{-1}(-{\bf  q}',-{\bf
q}).
\end{equation}

One can see that the Fourier transform of the electrochemical
potential in the Gaussian approximation is expressed in terms of the
structural factor of the ion subsystem and the Fourier transform of
the local part of the electron-ion interaction pseudopotential. The
time dependence is described by the average nonequilibrium value of
the electron density renormalized through the structural factor of
the ion subsystem, the pseudopotential $\omega_k({\bf  q})$, and the
function $\mathfrak M_{-k',-k}^{-1}(-{\bf  q}',-{\bf  q})$ inverse
to the pair irreducible cumulant average value of the electron
density fluctuation. Substituting (\ref{53}) into the expression for
the Massieu--Planck functional, we obtain
\[ \Phi^{(G)}(t)=\ln Z^G(t)=\beta\frac{N}{2S}\sum\limits_{
{\bf q}}{}^{'} \nu({\bf  q}|0)+\ln Z_{{\rm jell}}-\]
\[-\frac{1}{2}
 \sum_{k_1,k_2}\sum_{{\bf  q}_1,{\bf
q}_2}\langle\rho_{k_1}({\bf  q}_1)\rangle^t{\bar
S}_{k_{1}}^{-1}({\bf  q}_{1}) \times\]
\begin{equation}
\label{54}  \times\,{\mathfrak M}_{-k_1,-k_2}^{-1}(-{\bf  q}_1,-{\bf
q}_2){\bar S}_{k_{2}}^{-1}({\bf  q}_{2})\langle\rho_{k_2}({\bf
q}_2)\rangle^t.
\end{equation}
In this case, the entropy operator (\ref{51}) takes the form
\[\hat S^{(G)}(t)=\beta\frac{N}{2S}\sum\limits_{{\bf
q}}\,{}^{'}\nu({\bf  q}|0)+\ln
  Z_{{\rm
  jell}}-\]
\[-\frac{1}{2}\sum_{k_1,k_2}\sum_{{\bf  q}_1,{\bf  q}_2}\langle\rho_{k_1}
  ({\bf  q}_1)\rangle^t{\bar S}_{k_{1}}^{-1}({\bf  q}_{1}){\mathfrak M}_{-k_1,-k_2}^{-1}
  (-{\bf  q}_1,-{\bf  q}_2)\times\]
\[\times{\bar S}_{k_{2}}^{-1}({\bf  q}_{2})\langle\rho_{k_2}({\bf
q}_2)\rangle^t
   \!+\!\beta\Biggl(H\!-\!\frac{1}{SL}\sum_{k,{\bf  q}}\Big\{{\bar
   S}_k({\bf
   q})\!-\!\left(\frac{\beta}{SL}\right)^{-1}\!\!\!\!\times\]
   \begin{equation}\label{54a}
\times\,\sum_{k',{\bf  q}'}
   \langle\rho_{k'}({\bf  q}')\rangle^t{\bar S}_{k'}^{-1}({\bf  q}')
   {\mathfrak M}_{-k',-k}^{-1}(-{\bf  q}',-{\bf  q})\Big\}\rho_k({\bf  q})\!\Biggl).
\end{equation}
Instead of using the nonequilibrium statistical operator in the form
(\ref{1.23abb}) to obtain the transport equations for
$\langle\rho_k({\bf  q})\rangle^t$, we apply the
nonequilibrium statistical operator with regard for the projection,
which allows us to eliminate the time derivatives of thermodynamic
parameters \cite{l121,l122,Kost111}. We obtain
\[
\rho(t)= \rho_q(t)-\]
\begin{equation}\label{a38}
-\int\limits_{-\infty}^t \!\!{e}^{\varepsilon(t-t')}
T_q(t;t')\big(1-P_q(t')\big){i}L_N\rho(t'){\rm d}t',
\end{equation}
where
\[T_q(t,t')=\exp_+\left\{-\displaystyle\int\limits_{t'}^t
\big(1-P_q(t'')\big){i}L_N{\rm d}t''\right\}\] denotes the
generalized evolution operator with regard for projection and
$P_q(t')$ is the generalized Kawasaki--Gunton projection operator,
whose structure depends on the quasiequilibrium statistical
operator $\rho_q(t)$. In our case, $P_q(t)$ has the form
\[
 P_q(t)\rho'=\Bigg(\rho_q(t)-\sum_{k,{\bf  q}}
   \frac{\delta\rho_q(t)} {\delta\langle\rho_k({\bf  q})\rangle^t}
   \langle\rho_k({\bf  q})\rangle^t\Bigg)\textrm{Sp}\rho' + \]
   \begin{equation}\label{a39}
+\sum_{k,{\bf  q}}\frac{\delta\rho_q(t)} {\delta\langle\rho_k({\bf
q})\rangle^t}
   \textrm{Sp}\big(\rho_k({\bf  q})\rho'\big)
\end{equation}
and the operator properties: $P_q(t)\rho(t)=\rho_q(t),$
$P_q(t)\rho_q(t)=\rho_q(t),$ $P_q(t)P_q(t')=P_q(t).$ In order to
calculate the nonequilibrium statistical operator according to
(\ref{a38}) in the Gaussian approximation for $\rho^{(G)}_q(t)$, we must find,
first of all, the Kawasaki--Gunton projection operator.
Taking into account that
\[
\rho_q^{(G)}(t) = \exp\Bigg\{-\Bigg(\beta\frac{N}{2S}\sum_{\bf
q}{}'\nu({\bf  q}|0)+\ln Z_{{\rm jell}}-\]
\[-\frac{1}{2}\sum_{k_1,k_2}
\sum_{{\bf  q}_1,{\bf  q}_2}\langle\rho_{k_1}({\bf  q}_1)\rangle^t
{\bar S}_{k_{1}}^{-1}({\bf  q}_{1}) \times \]
\[\times\mathfrak M^{-1}_{-k_1,-k_2}(-{\bf  q}_1,-{\bf  q}_2){\bar S}_{k_{2}}^{-1}({\bf  q}_{2})
\langle\rho_{k_2}({\bf  q}_2)\rangle^t + \]
\[+\beta\Bigg(H-\frac{1}{SL}\sum_{k,{\bf  q}} \Bigg\{ {\bar
S}_k({\bf q})-\left(\frac{\beta}{SL}\right)^{-1}\!\! \sum_{k',{\bf
q}'}\langle\rho_{k'}({\bf  q}')\rangle^t \times\]
\begin{equation}\label{a40}
\times{\bar S}_{k'}^{-1}({\bf  q})\mathfrak M_{-k',-k}^{-1}(-{\bf
q}',-{\bf  q})\Bigg\}\rho_k({\bf  q}))\Bigg)\Bigg\}
\end{equation}
and
\[
\frac{\delta\rho_q^{(G)}(t)}{\delta\langle\rho_k({\bf
q}\,)\rangle^t}
       = -\sum_{k',{\bf  q}'}
      \big(\rho_{k'}({\bf  q}',\tau)- \]
\[ - \langle\rho_{k'}({\bf  q}')\rangle^t{\bar S}_{k\,'}^{-1}({\bf
q}' )\big)\mathfrak M_{-k',-k}^{-1}(-{\bf  q}',-{\bf q}\,){\bar
S}_{k}^{-1}({\bf  q})\rho_q^{(G)}(t),
\]
we obtain the following expression for the Kawasaki--Gunton
projection operator:
\[
P_q^{(G)}(t)\rho'=(\rho^{(G)}_q(t)+\]
\[ +\sum_{k,{\bf
q}}\sum_{k',{\bf  q}}
          \left(\rho_{k'}({\bf  q}',\tau)-\langle\rho_{k'}
     ({\bf  q}')\rangle^t{\bar S}_{k'}^{-1}({\bf  q}')\right)\times\]
\[\times{\mathfrak M}_{-k',-k}^{-1}(-{\bf  q}',-{\bf q})
     {\bar S}_{k}^{-1}({\bf  q})\langle\rho_k({\bf  q})\rangle^t\rho_q^{(G)}(t))
     \textrm{Sp}\rho'-\]
\[-\sum_{k,{\bf  q}'}\sum_{k',{\bf  q}'}
     \left(\rho_{k'} ({\bf  q}',\tau)-\langle\rho_{k'}({\bf  q}')\rangle^t
     {\bar S}_{k'}^{-1}({\bf  q}')\right)\times\]
     \begin{equation}\label{a41}
\times{\mathfrak M}_{-k',-k}^{-1}(-{\bf  q}',-{\bf  q}){\bar
S}_{k}^{-1}({\bf  q}) \textrm{Sp}(\rho_k({\bf
q})\rho')\rho_q^{(G)}(t).
\end{equation}
With regard for (\ref{a41}) and the relation
\[
 {i} L_N\rho_q(t) = \sum_{k,{\bf  q}}W^{(G)}(k,{\bf  q};t)\times \]
 \begin{equation}\label{a42}
\times\int\limits_0^1\!
 (\rho^{(G)}_{q})^{\tau}(t)
 \dot\rho_k({\bf  q})(\rho^{(G)}_{q})^{1-\tau}(t)\,{\rm  d}\tau,
\end{equation}
where
\[
W^{(G)}(k,{\bf  q};t)=\frac{\beta}{SL} \bar\mu_k({\bf  q};t)=\frac{\beta}{SL}\Bigg\{{\bar
S}_k({\bf  q})-\]
\begin{equation}\label{a43}
-\left(\frac{\beta}{SL}\right)^{-1}\!\!
 \sum_{k'{\bf  q}}
 \langle\rho_{k'}({\bf  q})\rangle^t{\bar S}_{k'}^{-1}({\bf  q})\mathfrak M_{-k',-k}^{-1}(-{\bf  q},-{\bf  q})\Bigg\},
\end{equation}
the nonequilibrium statistical operator can be written down in the
form
\[
  \rho(t)=\rho_q^{(G)}(t)-\sum_{k,\vec q}\int\limits_{-\infty}^t{\rm e}^{\varepsilon(t'-t)} T_q^G(t,t')\times
\]
\[
 \times\int\limits_0^1\! {\rm d}\tau(\rho^{(G)}_q(t'))^{\tau}I_\rho(k,\mathbf{q};t')\times
\]
\begin{equation}\label{a44}
\times(\rho^{(G)}_q(t'))^{1-\tau}
 W^{(G)}(k,\mathbf q;t'){\rm d}t',
\end{equation}
where
\begin{equation}\label{a45}
 I_\rho(k,\mathbf{q};t')=\big(1-{\cal P}^{(G)}(t')\big)i L_N \rho_k(\mathbf q)
\end{equation}
 is generalized diffusion flow,
${\cal P}^{(G)}(t)$ is projection operator acting on the operator
\begin{equation}\label{a46}
 {\cal P}^{(G)}(t)\hat{A}=\sum\limits_{k',\mathbf q'}\sum\limits_{k'',\mathbf q''}
 \delta\rho_{k'}(\mathbf q';t)\times
\end{equation}
\[
 \times{\mathfrak M}_{-k',-k''}(-\mathbf q',-\mathbf q'')\overline{S}_{k''}^{-1}(\mathbf q'')
 \langle\rho_{k''}(\mathbf q'')\widehat{A}\rangle_G^t,
\]
 where
 $\delta\rho_{k'}(\mathbf q';t)=\rho_{k'}(\mathbf q')-\langle\rho_{k'}(\mathbf q')\rangle^t\left(\frac{\beta}{SL}\right)^{-1}\overline{S}_{k'}^{-1}(\mathbf q')$,
$\langle(...)\rangle^{t'}_{(G)}={\rm Sp}(...\rho^{(G)}_q(t'))$ is
the averaging with the quasiequilibrium statistical operator in the
Gaussian approximation. In its structure, the nonequilibrium
statistical operator is a functional of the microscopic
fluxes $\dot\rho_k({\bf  q})$, the observable quantities
$\langle\rho_k({\bf  q})\rangle^t,$ and the quasiequilibrium
and equilibrium correlation functions of the electron subsystem of a
semibounded metal. With its help, we obtain the transport equation
for $\langle\rho_k({\bf  q})\rangle^t$ in the form
\[
  \frac{\partial}{\partial t}\langle\rho_k(\mathbf q)\rangle^t =
    \langle\dot\rho_k(\mathbf q)\rangle^t=
\]
\begin{equation}\label{a47}
  =-\sum_{k',\vec q'}\int\limits_{-\infty}^t\!\!\!{\rm e}^{\varepsilon(t'-t)}
    D^{(G)}_{JJ}(k,\mathbf q;k',\mathbf q';t,t')W^{(G)}(k',\mathbf q;t'){\rm d}t',
\end{equation}
where
\[
D^{(G)}_{JJ}(k,\mathbf q;k',\mathbf q';t,t')=
\]
\[=
  \left\langle I_\rho(k,\mathbf{q})T^G_q(t,t')
    I_\rho(k',\mathbf{q}';\tau)\right\rangle_{(G)}^{t'}=
\]
\[
=k\,\mathbf q \cdot \left\langle\vec J_k(\mathbf q)T_q^G(t,t')\mathbf
     J_{k'}(\mathbf q';\tau)\right\rangle_{(G)}^{t'}\cdot \mathbf q'k'=
\]
\begin{equation}\label{a48}
=k\,\mathbf q \cdot \tilde D^{(G)}_{JJ}(k,\mathbf q;k',\mathbf q';t,t')\cdot \mathbf q'k',
\end{equation}
 $\tilde D^{(G)}_{JJ}(k,{\bf  q};k',{\bf  q}';t,t')$ is the generalized diffusion
coefficient of electrons in a semibounded metal calculated using the quasiequilibrium
statistical operator in the Gaussian approximation.
%

\section{Approximation \boldmath$B_{k}({\bf q};t)B_{k'}({\bf q'};t)B_{k''}({\bf q''};t)$}

Let us consider the next approximation after the Gaussian one for the quasiequilibrium statistical sum
or the Massieu--Planck functional (\ref{551}). In this case, we obtain the following expression for the entropy operator:
\[
{\hat S}'(t)=\beta \frac {N}{2S} \sum_{{\bf q}}{}{}^{'}\nu({\bf
q}|0)+\ln Z_{{\rm jell}}-\]
\[ - \frac
{1}{2}\left(\frac{\beta}{SL}\right)^{2}
    \!\!\!\sum_{{\bf q}_{1},{\bf q}_{2}}\sum_{k_{1},k_{2}}B({\bf q}_{1},k_{1};t)
   B({\bf q}_{2},k_{2};t)\times\]
\[ \times  \mathfrak M_{-k_{1},-k_{2}}(-{\bf  q}_{1},-{\bf  q}_{2})+\]
\[ +\frac {i}{3!}\left(\frac{\beta}{SL}\right)^{3}\!\!\!
    \sum_{{\bf q}_{1},{\bf q}_{2},{\bf q}_{3}}
    \sum_{k_{1},k_{2},k_{3}}
    B({\bf q}_{1},k_{1};t)B({\bf q}_{2},k_{2};t) \times \]
\[ \times    B({\bf q}_{3},k_{3};t)
    {\mathfrak M}_{-k_{1},-k_{2},-k_{3}}(-{\bf  q}_{1},-{\bf
    q}_{2},-{\bf
    q}_{3})+\]
    \begin{equation}\label{nn020}
 +   \beta \Big(H-\frac{1}{SL}\sum_{{\bf q},k}{\bar\mu}_k({\bf
q};t)\rho_k({\bf  q})\Big).
\end{equation}
In order to eliminate the parameters ${\bar\mu}_k({\bf  q};t)$ from
this formula, the thermodynamic relation (\ref{n4*}) will be applied
once again:
\[\frac{\delta\Phi(t)}{\delta\frac{\beta}{SL}\bar\mu_k({\bf
q};t)}=\langle\rho_k({\bf  q})\rangle^t.\]
From here, we derive the equation for $\bar\mu_{k_i}({\bf  q}_1;t)$:
\[\langle\rho_k({\bf
q})\rangle^t=-\frac{\beta}{SL}\!\!
    \sum_{{\bf q}_{1},{\bf q}_{2}}
    \sum_{k_{1},k_{2}}\!
    \left(\!{\hat S}_{k_{1}}({\bf q}_{1})-{\bar \mu}_{k_{1}}({\bf
    q}_{1};t)\!\right)\!\times\]
\[ \times{\hat S}_{k_{2}}({\bf q}_{2})\mathfrak M_{-k_{1},-k_{2}}(-{\bf  q}_{1},-{\bf  q}_{2})
-\] \[ -\frac {i}{3!}\left(\frac{\beta}{SL}\right)^{2}\!\!\!
    \sum_{{\bf q}_{1},{\bf q}_{2},{\bf q}_{3}}\sum_{k_{1},k_{2},k_{3}}
    \left({\hat S}_{k_{1}}({\bf q}_{1})-{\bar \mu}_{k_{1}}({\bf q}_{1};t)\right)\times
    \]
\[ \times\left({\hat S}_{k_{2}}({\bf q}_{2})-{\bar \mu}_{k_{2}}({\bf
q}_{2};t)\right)
    {\hat S}_{k_{3}}({\bf q}_{3})\times \]
    \begin{equation}\label{nn120}
 \times   {\mathfrak M}_{-k_{1},-k_{2},-k_{3}}(-{\bf  q}_{1},-{\bf  q}_{2},-{\bf  q}_{3}).
\end{equation}
In its structure, this equation is square with respect to the
functions ${\bar \mu}_{k}({\bf q};t)$. To solve it approximately,
we replace one of the quantities ${\bar \mu}_{k_{i}}({\bf q}_{i};t)$ on the right-hand side of the
quadratic form by its value found in the Gaussian
approximation (\ref{53}). Then we obtain a linear equation for the
function ${\bar \mu}_{k}({\bf q};t)$:
\[
 \langle\rho_{k}({\bf  q})\rangle^t=
 -\frac{\beta}{SL}
 \sum_{k',{\bf  q}}\left({\bar S}_{k'}({\bf  q})-{\bar \mu}_{k'}({{\bf  q}}';t)\right)\times
 \]
 \begin{equation}\label{005}
\times G_{-k',-k}(-{{\bf  q}},-{{\bf  q}};t),
\end{equation}
where
\[ G_{-k_{1},-k_{2}}(-{\bf  q}_{1},-{\bf  q}_{2};t)=\]
\[ = {\bar S}_{k_{2}}({{\bf  q}}_{2}){\mathfrak M}_{-k_{1},-k_{2}}(-{{\bf
q}}_{1},-{{\bf q}}_{2})+\]
\[+ \frac{i}{2}\frac{\beta}{SL}\sum_{{\bf q}',{\bf
q}_{3}}\sum_{k',k_{3}}
      \langle\rho_{k'}({\bf  q}\,')\rangle^t {\bar
      S}_{k'}^{-1}({{\bf
      q}'})\times\]
\[\times {\mathfrak M}_{-k',-k_{2}}^{-1}(-{{\bf  q}'},-{\bf
q}_{2})\times\]
\begin{equation}\label{015}
 \times{\bar S}_{k_{3}}({{\bf  q}}_{3})
    {\mathfrak M}_{-k_{1},-k_{2},-k_{3}}(-{{\bf  q}}_{1},-{{\bf  q}}_{2},-{{\bf  q}}_{3}).
\end{equation}
As is seen, the function $G_{-k_{1},-k_{2}}(-{\bf  q}_{1},-{\bf
q}_{2};t)$ depends on time through the observable quantities
$\langle\rho_{k'}({\bf  q})\rangle^t$. It also depends on the
structural factor of the ion subsystem $S_{k'}({\bf  q})$, the
Fourier transform of the local part of the electron-ion interaction
pseudopotential $\omega_{k}({\bf  q})$, and the cumulant irreducible
average values of the electron density fluctuations: pair $\mathfrak
M_{-k',-k_{2}}(-{{\bf  q}},-{{\bf  q}}_{2})$ and triple ones
$\mathfrak M_{-k_{1},-k_{2},-k_{3}}(-{{\bf q}}_{1},-{{\bf
q}}_{2},-{{\bf  q}}_{3})$. The second term on the right-hand side of
Eq. (\ref{015}) involves the renormalization of the triple electron
correlations through the pair ones that make the dominant
contribution in the Gaussian approximation (see the previous
section).

Defining $G_{-k_{1},-k_{2}}^{-1}(-{\bf  q}_{1},-{\bf  q}_{2};t)$ as
a function inverse to $G_{-k_{1},-k_{2}}(-{\bf  q}_{1},-{\bf
q}_{2};t)$ by the relation
\[
  \sum_{k'',{\bf  q}''}G_{-k_{1},-k''}^{-1}(-{\bf  q}_{1},-{\bf  q}'';t)G_{-k'',-k_{2}}(-{\bf  q}'',-{\bf  q}_{2};t)=
  \]
\[ =\delta_{k_{1},k_{2}} \delta_{{\bf  q}_{1},{\bf  q}_{2}}
 \]
and using Eq.(\ref{005}), we derive the following expression for the
Fourier transform of the electron electrochemical potential:
\[
 {\bar \mu}_k({\bf  q};t)= {\bar S}_{k}({\bf  q})-\]
 \begin{equation}\label{053}-
 \left(\frac{\beta}{SL}\right)^{-1}\!\!
 \sum_{k',{\bf  q}}
 \langle\rho_{k'}({{\bf  q}}\,')\rangle^t G_{-k',-k}^{-1}
 (-{{\bf  q}}\,',-{{\bf  q}};t).
\end{equation}
Now, with regard for (\ref{053}), the entropy operator (\ref{nn020}) can
be presented in the form
\[{\hat S}'(t)=\beta\frac{N}{2S}\sum_q{}'\nu(q|0)+\ln Z_{{\rm jell}}-
\]
\[-\frac{1}{2}
     \sum_{k',k''}\!\!\sum_{{\bf q}',{\bf q}''}\!\!\langle\rho_{k'}({\bf q})\rangle^t
     {\bar G}_{k',k''}^{(2)}({\bf q}',{\bf q}'';t)
     \langle\rho_{k''}({\bf q}'')\rangle^t\!+\]
\[+\frac{i}{3!}
              \sum_{k',k'',k'''}\sum_{{\bf q}',{\bf q}'',{\bf q}'''}
  {\bar G}_{k',k'',k'''}^{(3)}({\bf q}',{\bf q}'',{\bf
  q}''';t)\times\]
\[\times\left\langle\rho_{k'}({\bf q}')\right\rangle^t
  \left\langle\rho_{k''}({\bf q}'')\right\rangle^t
  \left\langle\rho_{k'''}({\bf q}''')\right\rangle^t+  \]
\[
+\beta\Bigg(H-\frac{1}{SL}\sum_{k,{\bf q}}\Bigg(\bar S_k({\bf
q})-\left(\frac{\beta}{SL}\right)^{-1}
  \sum_{k',{\bf q}'}\langle\rho_{k'}({\bf q}')\rangle^t\times \]
  \begin{equation}\label{054}
 \times G^{-1}_{k'k}({\bf q},{\bf q}';t)\Bigg)\rho_k({\bf q})\Bigg),
\end{equation}
where
\[
 \bar {G}_{k',k''}^{(2)}({\bf q}',{\bf q}'';t)=
    \sum_{k_{1},k_{2}}\sum_{{\bf q}_{1},{\bf q}_{2}}
    G_{k',k_{1}}^{-1}({{\bf  q}}\,\,',{{\bf  q}}_{1};t) \times \]
 \[ \times{\mathfrak M}_{-k_{1},-k_{2}}(-{{\bf  q}}_{1},-{{\bf  q}}_{2})G_{k_{2},k''}^{-1}({{\bf  q}}_{2},{{\bf  q}}\,\,'';t),
\]
\[
 {\bar G}_{k',k'',k'''}^{(3)}({\bf q}',{\bf q}'',{\bf q}''';t)= \]
\[ = \sum_{k_{1},k_{2},k_{3}}\sum_{{\bf q}_{1},{\bf q}_{2},{\bf
q}_{3}} G_{k',k_{1}}^{-1}({{\bf  q}}',{{\bf  q}}_{1};t)\times   \]
\[ \times G_{k'',k_{2}}^{-1}({{\bf  q}}\,'',{{\bf  q}}_{2};t)G_{k''',k_{3}}^{-1}({{\bf
q}}\,''',{\bf q}_{3};t)\times\]
\[ \times {\mathfrak
M}_{-k_{1},-k_{2},-k_{3}}(-{{\bf  q}}_{1},-{{\bf  q}}_{2},-{\bf
q}_{3}).
\]

These functions involve the dynamic renormalization of the
cumulant irreducible average values of the electron density
fluctuations: pair $\mathfrak M_{-k',-k_{2}}(-{{\bf  q}},-{\bf
q}_{2})$ and triple ones $\mathfrak M_{-k_{1},-k_{2},-k_{3}}(-{\bf
q}_{1},-{{\bf  q}}_{2},-{{\bf  q}}_{3})$ through functions
(\ref{015}). With regard for the entropy operator (\ref{nn020}), the
quasiequilibrium statistical operator reads
 \[\rho^{(G+1)}_q(t)=\exp\Bigg(-\Big(\beta\frac{N}{2S}\sum_{\bf q}{}'
            \nu({\bf q}|0)+\ln Z_{{\rm jell}}-\]
\[-\frac{1}{2}
     \sum_{k',k''}\sum_{{\bf q}',{\bf q}''}
     \langle\rho_{k'}({\bf q})\rangle^t\bar{G}_{k',k''}^{(2)}({\bf q}',{\bf q}'';t)
     \times\]
\[\times     \langle\rho_{k''}({\bf q}'')\rangle^t+  \]
\[+\frac{i}{3!}
                 \sum_{k',k'',k'''}\sum_{{\bf q}',{\bf q}'',{\bf q}'''}
                \bar{G}_{k',k'',k'''}^{(3)}({\bf q}',{\bf q}'',{\bf
                q}''';t)\times\]
\[\times\langle\rho_{k'}({\bf q}')\rangle^t
\langle\rho_{k''}({\bf q}'')\rangle^t\langle\rho_{k'''}({\bf
q}''')\rangle^t+
\]
\[+\beta\Bigg(H-\frac{1}{SL}\sum_{k,{\bf q}}\Big(\bar S_k({\bf
q})-\Big(\frac{\beta}{SL}\Big)^{-1}\!\!
  \sum_{k',{\bf q}'}\langle\rho_{k'}({\bf q}')\rangle^t \times \]
  \begin{equation}\label{055}
\times  G^{-1}_{k'k}({\bf q},{\bf q}';t))\rho_k({\bf q})\Big)\Bigg),
\end{equation}
where the index ``$(G+1)$'' denotes the third order with respect to
the observable parameters in the quasiequilibrium statistical
operator. In addition to the pair (Gaussian) one, it also allows for
the cubic dependence on the parameters of the reduced description
$\langle\rho_{k}({\bf q})\rangle^t$ with dynamic renormalizations in
the functions $\bar {G}_{k',k''}^{(2)}({\bf q}',{\bf q}'';t)$ and
${\bar G}_{k',k'',k'''}^{(3)}({\bf q}',{\bf q}'',{\bf q}''';t)$.
Since
\[
{i}L_N\rho^{(G+1)}_q(t)=\sum_{k,{\bf  q}}W^{(G+1)}(k,{\bf
q};t)\times\]
\begin{equation}\label{a420}
\times\int\limits_0^1\!(\rho^{(G+1)}_{q})^{\tau}(t)\dot\rho_k({\bf
q})(\rho^{(G+1)}_{q})^{1-\tau}(t) {\rm d}\tau,
\end{equation}
where
\[
W^{(G+1)}(k,{\bf  q};t)=\frac{\beta}{SL} \bar\mu_k({\bf  q};t)
  =\frac{\beta}{SL}\Bigg\{{\bar S}_k({\bf  q})-\]
\begin{equation}\label{a430}
-\left(\frac{\beta}{SL}\right)^{-1}\sum_{k'{\bf  q}}
\langle\rho_{k'}({\bf  q})\rangle^t G_{-k',-k}^{-1}(-{\bf q}',-{{\bf
q}};t)\Bigg\},
\end{equation}
we obtain the following expression for the nonequilibrium statistical operator in approximation (\ref{055}):
\[
\rho(t)=\rho^{(G+1)}_q(t)-
  \sum_{k,{\bf  q}}\int\limits_{-\infty}^t\!{e}^{\varepsilon(t'-t)}
  T^{(G+1)}_q(t,t')\times\]
\[\times\big(1-P^{(G+1)}_q(t')\big)\int\limits_0^1\!(\rho^{(G+1)}_{q})^{\tau}(t')
    \dot\rho_k({\bf  q})(\rho^{(G+1)}_{q})^{1-\tau}(t')\times\]
    \begin{equation}\label{056}
\times W^{(G+1)}(k,{\bf  q};t'){\rm d}t',
\end{equation}
where the Kawasaki--Gunton projection operator $P^{(G+1)}_q(t')$ and
the respective evolution operator $T^{(G+1)}_q(t,t')$ are calculated
with the quasiequilibrium statistical operator in approximation
(\ref{055}). Moreover, $P^{(G+1)}_q(t')$ has the following structure:
\[ P_q^{(G+1)}(t)\rho'=\Bigg(\rho^{(G+1)}_q(t)-
\]
\[
-\sum_{k,{\bf  q}}\Bigg\{
       \Big(\beta \rho_{k'}({\bf q}',\tau)G^{-1}_{k'k}({\bf q},{\bf
q}';t)-          \]
\[ - \langle\rho_{k'}({\bf q}')\rangle^t \bar
{G}_{k',k}^{(2)}({\bf q}',{\bf q};t)\Big)\times\]
\[ \times\frac
{i}{2}\sum_{k',k''}\sum_{{\bf
q}',{\bf q}''}\langle\rho_{k'}({\bf q}')\rangle^t
\langle\rho_{k''}({\bf q}'')\rangle^t\times\]
\[ \times \bar{G}_{k',k'',k}^{(3)}({\bf q}',{\bf q}'',{\bf q};t)\}\langle\rho_{k}({\bf q})\rangle^t \rho^{(G+1)}_q(t)){\rm Sp}
(\rho')\times\] \[+\sum_{k,{\bf
q}}\Bigg\{(\beta
\rho_{k'}({\bf q}',\tau)G^{-1}_{k'k}({\bf q},{\bf q}';t)- \]
\[-\langle\rho_{k'}({\bf q}')\rangle^t
\bar {G}_{k',k}^{(2)}({\bf q}',{\bf q};t))\times\] \[\times \frac
{i}{2}\sum_{k',k''}\sum_{{\bf
q}',{\bf q}''}\langle\rho_{k'}({\bf q}')\rangle^t
\langle\rho_{k''}({\bf q}'')\rangle^t\times\]
\begin{equation}\label{057}
\times\bar{G}_{k',k'',k}^{(3)}({\bf q}',{\bf q}'',{\bf q};t)\Bigg\}
 {\rm Sp} \big( \rho_{k}({\bf q}) \rho'\big)\rho^{(G+1)}_q(t).
\end{equation}

As compared to the result of action of the operator $P_q^{(G)}(t)$
in the Gaussian approximation, it already includes the third order
with respect to the parameters of the reduced description. With the
help of the nonequilibrium statistical operator (\ref{055}), we obtain
the transport equation for $\langle\rho_k({\bf  q})\rangle^t$ in the
form
\[
\frac{\partial}{\partial t}\langle\rho_k({\bf  q})\rangle^t
  =\langle\dot\rho_k({\bf  q})\rangle_{(G+1)}^t-\]
\[-\sum_{k',{\bf  q}'}
   \int\limits_{-\infty}^t\!\!{e}^{\varepsilon(t'-t)}D^{(G+1)}_{JJ}(k,{\bf  q};k',{\bf  q}';t,t')\times
   \]
\[ \times   W^{(G+1)}(k',{\bf  q};t'){\rm d}t'+\]
\[+\sum_{k',{\bf
q}'}\int\limits_{-\infty}^t{e}^{\varepsilon(t'-t)}
            \langle\dot\rho_k({\bf  q})T_q^{(G)}(t,t')P_q^{(G+1)}(t')\times
            \]
            \begin{equation}\label{058}
 \times  \dot\rho_{k'}({\bf
q}\,';\tau)\rangle_{(G+1)}^{t'}W^{(G+1)}(k',{\bf  q}';t')\,{\rm d}t',
\end{equation}
where
\[
D^{(G+1)}_{JJ}(k,{\bf  q};k',{\bf  q}';t,t')=\]
 \[ =\langle\dot\rho_k({\bf  q})T^G_q(t,t')\dot\rho_{k'}({\bf
 q}\,',\tau)\rangle_{(G+1)}^{t'}=\]
\[=k\,{\bf  q} \,\langle{\bf  J}_k({\bf  q})T_q^G(t,t'){\bf  J}_{k'}
 ({\bf  q}';\tau)\rangle_{(G+1)}^{t'}\, {\bf  q}'\,k'=\]
 \begin{equation}\label{059}
=k\,{\bf  q} \, \tilde D^{(G+1)}_{JJ}(k,{\bf  q};k',{\bf q}\,';t,t')\,
{\bf  q}'\,k',
\end{equation}
$\tilde D^{(G+1)}_{JJ}(k,{\bf  q};k',{\bf  q}';t,t')$ is the
generalized diffusion coefficient of electrons in a semibounded
metal calculated with the quasiequilibrium statistical operator in
approximation (\ref{055}). With regard for the structure of
functions (\ref{015}) and (\ref{a430}) and the Kawasaki--Gunton projection
operator (\ref{057}), we can conclude that Eq. (\ref{058}) is
nonlinear with respect to $\langle\rho_k({\bf  q})\rangle^t$.

\section{Conclusions}

Electrodiffusion processes in the electron subsystem of a
semibounded metal are described on the basis of the generalized
``jellium'' model with the use of the NSO method, where the only
parameter of the reduced description is the nonequilibrium average
value of the electron density. Applying the functional integration
technique, we have calculated the quasiequilibrium statistical sum for
such a system in the case of the local pseudopotential of electron-ion
interaction in a metal in the Gaussian and higher approximations
with respect to the dynamic electron correlations. They are used
to obtain expressions for the nonequilibrium statistical
operator in the Gaussian and higher approximations with respect to
the dynamic electron correlations, which makes it possible to go
beyond the linear approximation with respect to the electrochemical
potential. In the respective approximations for the nonequilibrium
statistical operator, we have derived the generalized transport equations
(generalized diffusion equations) for the nonequilibrium average
value of the electron density that can be applied to the description
of strongly nonequilibrium processes for the electron subsystem of a
semibounded metal. The generalized diffusion coefficients for
electrons in a semibounded metal that enter the corresponding
transport equations are calculated with the quasiequilibrium
statistical operator in the respective approximations: Gaussian one
(\ref{a40}) and approximation (\ref{055}). An important point in such an approach
is that the time correlation functions and the generalized diffusion
coefficients are calculated with the quasiequilibrium statistical
operator in the corresponding approximation and represent
functionals of the observable quantities $\langle\rho_k({\bf
q})\rangle^t$ of a certain order. Of special interest in this approach
are the investigations of the dynamic structural factor for the
nonequilibrium electron subsystem of a semibounded metal.

\rezume{%
ДО СТАТИСТИЧНОГО ОПИСУ ЕЛЕКТРОДИФУЗІЙНИХ\\ ПРОЦЕСІВ ЕЛЕКТРОННОЇ
ПІДСИСТЕМИ\\ НАПІВОБМЕЖЕНОГО~~~~~~~ МЕТАЛУ\\ В УЗАГАЛЬНЕНІЙ МОДЕЛІ
``ЖЕЛЕ''}{П.П. Костробій, Б.М. Маркович, А.І. Василенко, \\М.В.
Токарчук} {За допомогою методу функціонального інтегрування отримано
 нерівноважний статистичний оператор для електронної підсистеми
 напівобмеженого металу в узагальненій моделі  ``желе'' у гаусовому  та
 вищих наближеннях\rule{0pt}{13pt} за динамічними електронними кореляціями при розрахунку\rule{0pt}{13pt}
 квазірівноважної статистичної суми.
 Такий підхід дає можливість\rule{0pt}{13pt} вийти за межі лінійного наближення за градієнтом
 електрохімічного потенціалу,\rule{0pt}{13pt} яке відповідає слабо нерівноважним процесам, та отримати\rule{0pt}{13pt}
 узагальнені\rule{0pt}{13pt} рівняння переносу, які описують нелінійні\linebreak процеси.\rule{0pt}{13pt}}

\end{document}